# Autaptic Connections Shift Network Excitability and Bursting

Laura Wiles[1], Shi Gu[1,2], Fabio Pasqualetti[3], Danielle S. Bassett[1,4],

David F. Meaney[1,5]

[1] Department of Bioengineering, School of Engineering and Applied Sciences, University of Pennsylvania, Philadelphia, PA 19104 USA

[2] Applied Mathematics and Computational Science Graduate Program, School of Arts and Sciences, University of Pennsylvania, Philadelphia, PA 19104 USA

[3] Department of Mechanical Engineering, University of California Riverside, CA USA

[4] Department of Electrical and Systems Engineering, School of Engineering and Applied Sciences, University of Pennsylvania, Philadelphia, PA 19104 USA

[5] Department of Neurosurgery, Perelman School of Medicine, University of Pennsylvania, Philadelphia, PA 19104 USA

# Abstract

Network architecture forms a critical constraint on neuronal function. Here we examine the role of structural autapses, when a neuron synapses onto itself, in driving network-wide bursting behavior. Using a simple spiking model of neuronal activity, we study how autaptic connections affect activity patterns, and evaluate if neuronal degree or controllability are significant factors that affect changes in bursting from these autaptic connections. We observed that adding increasing numbers of autaptic connections to excitatory neurons increased the number of spiking events in the network and the number of network-wide bursts, particularly in the portion of the phase space in which excitatory synapses were stronger contributors to bursting behavior than inhibitory synapses. In comparison, autaptic connections to excitatory neurons with high average controllability led to higher burst frequencies than adding the same number of self-looping connections to neurons with high modal controllability. The number of autaptic connections required to induce bursting behavior could be lowered by selectively adding autapses to high degree excitatory neurons. These results suggest a role of autaptic connections in controlling network-wide bursts in diverse cortical and subcortical regions of mammalian brain. Moreover, they open up new avenues for the study of dynamic neurophysiological correlates of structural controllability.

# Introduction

Network architecture forms a critical driver for complex function in a broad class of systems across biomedical science [1-3]. Transcription networks play a critical role in the nascent field of synthetic biology [4], while protein-signaling networks offer an unparalleled opportunity for drug discovery by solving the challenging problem of combinatorial therapies [5]. Similarly, understanding gene regulatory networks has proven crucial for identifying classes of patients that may respond best to particular therapies, supporting the emerging field of personalized medicine [6]. For the brain, transcribing the architecture of white matter tracts crisscrossing the human cortex [7, 8] have offered inherently new ways to explain the relationship between brain and behavior [9, 10], and its alteration in neurological disorders and psychiatric disease [11-14]. In each of these contexts, the organization of connectivity patterns plays a key role in constraining system dynamics and organism function [15].

Despite the excitement engendered by these results across a wide variety of fields, groundbreaking new discoveries will necessitate a fundamental turn from descriptive statistics towards answering questions to identify the network features that control the temporal dynamics of these networks [16, 17]. For example, what are the mechanisms by which network structure affects functional dynamics? And how could one intervene in a network to push the system dynamics towards a new state? Recent advances from the emerging field of network control theory [18, 19] offer conceptual frameworks and mathematical tools to address exactly these questions, thereby promising progress on an entirely new class of problems. Based on the architecture of the network, one can identify points (nodes) within the network that offer the most energetically effective means for controlling network dynamics towards specific states [18]. These concepts and tools have direct implications for engineering system robustness (networks resistant to control), sensitivity (networks amenable to control), and specificity (networks resistant to some control strategies but not others).

Control input to a network is traditionally formalized as a scalar value added to a node's state, representing an external intervention to alter the dynamics of individual nodes, and by extension distributed networks [20-22]. Intuitive examples include electrical stimulation to nodes in an electrical or biological circuit, mechanical perturbations to points in a material, or magnetic modulation to regions of a patient's brain [16]. Yet, these notions of controlling dynamics span only a narrow set of the broader class of control strategies potentially accessible to a biological system [23, 24]. Indeed, control can more generally be conceptualized as a change to a node's dynamics that is achieved by altering the physical structure of that node, or by altering the dynamic processes occurring on top of that structure.

A quintessential example of this broader notion of control is that of self-loops [25]. A self-loop is either an abstract or physical connection that a node makes onto itself. The presence of self-loops can fundamentally change the intrinsic dynamics of the nodes at which they occur. Yet how these changes may affect the dynamics of the distributed network is less clear. How might self-loops be used by biological systems to control systemic dynamics? And where should such loops be placed in the network to affect a specific dynamic outcome? Moreover, how do these rules depend on the network structure and initial conditions of the system?

These questions are particularly meaningful in neural networks, where self-looping structures are known as "autapses". An autapse is a synaptic connection from a neuron onto itself, either across separate dendrites of the same neurons or from the dendritic arbor to the parent cell body. Since their discovery over four decades ago, autapses are now documented in pyramidal neurons within the developing rat neocortex [26] and the cat visual cortex [27], appear more commonly on inhibitory neurons [28-30], and appear abundantly in fast-spiking interneurons, but not in low-threshold spiking interneurons [29]. In other cases, autaptic connections can represent only a small number of the thousands of excitatory and inhibitory synaptic connections received by a neuron [31], yet their self-stimulating nature can provide a very economical method to affect neuronal activity dynamics. To this end, several studies show that the delays in autaptic inputs affect the bursting behavior and information transfer of individual neurons, offering insights into regulating the activity of these neurons [32-35]. However, relatively little is known on the consequences of self-loop connections at the network scale, and how these connections affect the overall dynamics of the neural network. As such, principles explaining the functional network role of autapses in neural circuits remain a mystery.

In this communication, we study self-looping in cortical and hippocampal neuronal networks and examine the impact of these loops on activity dynamics, from network-wide bursting to coordinated firing of neuronal subpopulations. We test the specific situation in which self-loops are placed either randomly throughout the network or at driver nodes predicted to facilitate different control strategies. We advance several new findings that show that autaptic connections enhance the network's excitability, increasing bursting frequency and regularity. For the networks studied, effects of autaptic connections are strongest when these connections are placed on excitatory neurons; when the number of autaptic connections is held constant, bursting frequency is higher when more autapses are placed on fewer neurons rather than when fewer autapses are placed on more neurons. Finally, we observed the greatest increase in network-wide bursting when autapses were located at points in the network that are theoretically predicted to be effective controllers.

## Methods

To study the relationship between structural connectivity and neuronal network dynamics, we constructed computational neural networks and simulated their activity using Izhikevich integrate and fire model neurons [36, 37]. Preliminary simulations showed that network activity dynamics did not change for networks containing more than 800 neurons. We therefore used a network size of 1000 neurons for all simulations. All simulations were completed using in-house software developed in the MATLAB programming language (MathWorks, Inc.).

**Neural Network Simulations: Neurons and Their Coupling**

In each simulation, we placed 800 excitatory and 200 inhibitory neurons uniformly at random on the surface of a unit sphere using MATLAB's twister random number generator. The number of outputs for each neuron was generated from a normal distribution with a mean of 93.75 outputs per neuron and a variance of 9.375, resulting in a mean of 75 excitatory inputs and 18.75 inhibitory inputs per neuron. We chose these values to reflect anatomical estimates from empirical data that suggest that – in cortex – approximately 20% of neuron inputs are inhibitory and approximately 20% of neurons themselves are inhibitory [38]. We placed these output and input connections within the network in a distance-dependent manner, consistent with prior empirical studies [39, 40].

We connected outputs from each neuron to other neurons using a distance-dependent drop-off probability function $P_{ij} = 1/d^2$, where $d$ is the arc length between node $i$ and node $j$ along the surface of the sphere. Collectively, these connections between all possible pairs of nodes formed the connectivity matrix, $A$. The weight of an edge, codified in the element $A_{ij}$, represents an aggregate synaptic strength drawn from a normal distribution with a specified mean strength and a standard deviation of 0.1, consistent with prior work [41-44]. We used a standard deviation of 0.1 to maximize variance of connection weights while minimizing overlap among synaptic strengths from simulations with different mean strengths. For example, if we have two different networks with mean excitatory strengths of 2 and 3, with a standard deviation of 0.1, nearly all of the individual strengths will fall in the ranges of 1.7-2.3 and 2.7-3.3, respectively, which allows for a distribution of synaptic weights while maintaining the difference between these two networks. The diagonal elements of the weighted adjacency matrix $A$ are equal to zero, representing the fact that there are no self-connections (or autapses) present.

**Neural Network Simulations: Model of Dynamics**

We model neuronal activity with systems of ordinary differential equations, following the work of Izhikevich in 2003. First, we define the neuron's membrane potential, membrane recovery variable, and after-spike reset values as follows:

$$v' = 0.04v^2 + 5v + 140 - u + I, \qquad [1]$$

$$u' = a(bv - u), \qquad [2]$$

$$\text{if } v \geq 30 \text{ mV, then } \begin{cases} v \leftarrow c \\ u \leftarrow u + d \end{cases},$$

where the dimensionless variable $v$ represents the neuron's membrane potential and $u$ represents the neuron's membrane recovery variable. Each neuron is assigned the parameters $a$, $b$, $c$, and $d$, which govern the intrinsic behaviors and dynamics of the neurons [36]. For our simulations, we assigned values for $a$, $b$, $c$, and $d$ such that the behavior of excitatory neurons would be characterized by regular-spiking, consistent with the majority of neurons in the cortex, but still exhibit enough heterogeneity that any two neurons would never display identical dynamics [36]. For inhibitory neurons, values for $a$, $b$, $c$, and $d$ were assigned such that both fast-spiking and low-threshold spiking interneurons existed in the simulated system [36].

Following [37], we applied a random thalamic input to the network of 1 Hz, consistent with the mean firing rates of cortical neurons observed *in vivo* [45, 46]. We included the exponential decay of synaptic currents. The rate of membrane potential change was capped (225 mV/ms) to avoid unrealistic membrane potentials (> 50mV) during a spiking event. When a neuron fired an action potential, the current was injected into output neurons in the next time step (.2 ms later).

**Normative Dynamics**

We studied neuronal dynamics in networks characterized by different excitatory and inhibitory strengths to identify excitation and inhibition levels that produced similar spiking activity. Intuitively, at different excitation and inhibition levels, a neuron might require fewer or more synchronous inputs to fire an action potential. We examined 10 mean excitatory strengths, from 1 to 10 in unit increments. With a mean excitatory connection strength of 1, a neuron would need to receive approximately 20 synchronous inputs to fire an action potential; with an excitatory connection strength of 10, a neuron would only need to receive two synchronous inputs to fire an action potential [37]. To complement these 10 excitation levels, we also examined 10 mean inhibitory strengths, from -10 to -1 in increments of unity. To achieve numerical stability and obtain robust results, we performed ten 120s stimulations with 5 steps/ms for each possible combination of mean excitatory strength and mean inhibitory strength. We then analyzed the resultant spiking behavior to measure firing rate and to isolate bursts.

**Addition of Autapses**

We added autapses, defined as self-loops in the network (mathematically: non-zero elements on the diagonal of the connectivity matrix $A$), to either excitatory or inhibitory neurons according to two characteristics of that node's connections: strength and controllability (Figure 1C, D). Strength is defined as the sum of the inputs onto and outputs from that neuron. Controllability can be separated into notions of average control and modal control, which are defined in detail in the next section. Here we simply describe these notions intuitively. Average control describes the theoretically predicted preference for the node to push the system into local easily-reachable states, and modal control describes the theoretically predicted preference for the node to push the system into distant difficult-to-reach states. Strength, average control, and modal control provide complementary estimates of the influence a node has on network dynamics.

When adding autapses, we implemented seven targeting strategies, adding autapses to neurons with the (1) highest strength, (2) lowest strength, (3) highest average controllability, (4) lowest average controllability, (5) highest modal controllability, and (6) lowest modal controllability as well as (7) neurons chosen uniformly at random. To construct appropriate null models for our subsequent analyses, we consider the fact that by adding autapses to a neuron, we increased both the number of output connections from and the number of input connections to the selected neuron. We therefore implemented two null models. First, we constructed a null model that accounts for the increase in outputs on autaptic neurons by adding outputs from the would-be autaptic neurons to other neurons in the network in a distance-dependent manner. Second, we constructed a null model that accounts for the increase in inputs on autaptic neurons by adding inputs from other neurons to the would-be autaptic neuron.

In both autaptic network and null models, we added connections (self-loops or non-selfloops, respectively) to between 10% and 100% (in increments of 10%) of either excitatory or inhibitory neurons. Given the relatively rare frequency of autaptic connections in vivo, we added small amounts of autaptic or non-autaptic connections (1%, 2%, 3%, 4%, 5%,) to the selected neurons. Autaptic connections were added as self-looping connections in proportion to the outputs from a given neuron; e.g., a neuron with 100 separate outputs to other neurons received 3 additional self-looping (autaptic) connections to produce 3% new autaptic connections. Non-autaptic connections followed a similar mapping. To maintain consistency with previous simulations, current was injected from these autaptic connections at the next timestep. To obtain robust results, we completed ten sets of simulations for each combination of excitatory and inhibitory strengths. We performed each of the ten simulations at a given excitation and inhibition level on different original connectivity matrix with no autapses, which we then modified by adding autaptic or non-autaptic connections according to the targeting strategies described above. From each modified network, we analyzed spiking activity to better understand the effects of targeted connectivity changes on bursting behavior.

**Targeting Strategies**

We employ the targeted addition of autapses to neurons as a means to study potential mechanisms by which a network can control its global dynamics. The simplest notion of a node that has a high level of influence on dynamics is the notion of a hub [47]. A hub is a node that has either many connections emanating from it (high degree), or on average very strong connections emanating from it (high strength) or both. Here because we are studying inherently weighted graphs, we study a neuron's strength, defined as the sum of the inputs onto and outputs from that neuron. In prior studies, this metric of influence has been shown to be an indirect proxy for controllability [16, 48] and to correlate with statistical measurements of system dynamics [49-51].

Arguably a more direct measure of influence is one that would take into account not just which connections a neuron had, but also how the neuron used them. Philosophically, the notion of *influence* is essentially a *dynamical* notion, implying change in a system's state. Thus, for a more direct measure of influence, we turned to applications of dynamical systems theory to the problem of network control. In network control theory, one wishes to understand how to drive a networked system from a specified initial state to a specified target state in finite time and with limited energy. The rather nascent field has developed a theoretical framework, analytical results, and statistical tools that can be used to identify control points, which are theoretically predicted to be critical for driving the network's observed dynamics [18, 19].

Traditionally utilized to study technological, robotic, and mechanical systems, network control theory offers a particularly appealing conceptual and mathematical framework in which to study neural systems [16, 52, 53]. In this context, control points in the network are neurons that are predicted to be critical for driving large-scale neural dynamics. To identify these control neurons, we implement a linearized generalization of nonlinear models of cortical activity [54, 55]. Specifically, we used a noise-free linear discrete-time and time-invariant model of network dynamics [16],

$$\mathbf{x}(t+1) = \mathbf{A}\mathbf{x}(t) + \mathbf{B}_\kappa \mathbf{u}_\kappa(t), \qquad [3]$$

where $\mathbf{x}$ describes the state of neurons over time, and $\mathbf{A}$ is a signed, weighted, directed adjacency matrix whose elements, $A_{ij}$, specify the strength of the connection from node $i$ to node $j$ (after a normalization to ensure Schur stability). The matrix $\mathbf{B}_\kappa$ is an input matrix that identifies the control neurons, $\kappa = \{k_1, ..., k_m\}$, such that $B_\kappa = [e_{k1} ... e_{km}]$, where $e_i$ notes the $i$-th canonical vector of dimension N, and N is the number of neurons in the network. The signal $u_\kappa$ is the control input to the control neurons.

Using this model, we can define two distinct controllability strategies: the average controllability and the modal controllability, which – as mentioned earlier – describe the ability to push a system into local easily-reachable states or into distant difficult-to-reach states, respectively. To define the notion of average controllability, we first write down the controllability Gramian, $\mathbf{W}_\kappa$, as

$$\mathbf{W}_\kappa = \sum_{\tau=0}^{\infty} \mathbf{A}^\tau \mathbf{B}_\kappa \mathbf{B}_\kappa^T (\mathbf{A}^T)^\tau, \qquad [4]$$

where T indicates a matrix transpose and τ is a constant ranging from 0 to infinity. Then, average controllability is defined as the trace of the inverse of the controllability Gramian Trace($\mathbf{W}_\kappa^{-1}$), but for computational purposes can also be approximated via Trace($\mathbf{W}_\kappa^{-1}$) (see [16]). Thus, to identify nodes with the highest average controllability, we select nodes that maximize Trace($\mathbf{W}_\kappa$). Since the trace is a linear mapping and is invariant under cyclic permutations, we note that

$$\mathrm{Tr}(\mathbf{W}_\kappa) = \mathrm{Tr}(\sum_{\tau=0}^{\infty} \mathbf{A}^\tau \mathbf{B}_\kappa \mathbf{B}_\kappa^T (\mathbf{A}^T)^\tau) = (\sum_{\tau=0}^{\infty} \mathrm{Tr}(\mathbf{A}^\tau \mathbf{B}_\kappa \mathbf{B}_\kappa^T (\mathbf{A}^T)^\tau)) =$$
$$(\sum_{\tau=0}^{\infty} \mathrm{Tr}(\mathbf{B}_\kappa \mathbf{B}_\kappa^T (\mathbf{A}^T)^\tau \mathbf{A}^\tau)) = \mathrm{Tr}(\mathbf{B}_\kappa \mathbf{B}_\kappa^T \sum_{\tau=0}^{\infty} (\mathbf{A}^T)^\tau \mathbf{A}^\tau) = \sum_{i \in \kappa} (\sum_{\tau=0}^{\infty} (\mathbf{A}^T)^\tau \mathbf{A}^\tau)_{ii}, \qquad [5]$$

where $(\sum_{\tau=0}^{\infty}(\mathbf{A}^T)^\tau \mathbf{A}^\tau)_{ii}$ represents the *i*-th diagonal entry of the matrix $\sum_{\tau=0}^{\infty}(\mathbf{A}^T)^\tau \mathbf{A}^\tau$. To maximize the trace, we chose the set of nodes, κ, containing the largest diagonal entries of $\sum_{\tau=0}^{\infty}(\mathbf{A}^T)^\tau \mathbf{A}^\tau$. If **A** is stable, then $X = \sum_{\tau=0}^{\infty}(\mathbf{A}^T)^\tau \mathbf{A}^\tau$ is the solution to the discrete-time Lyapunv equation, $AXA^T - X + Q = 0$, where $Q = \mathbf{B}_\kappa \mathbf{B}_\kappa^T$. We then assign a ranked value of average controllability between 1 and N to each neuron, with 1 representing the neuron with the lowest average controllability and N representing the neuron with the highest average controllability.

To complement the notion of average control, we also define modal controllability, which is highest in nodes that can steer the system toward difficult-to-reach states. Modal controllability is calculated from the eigenvector matrix $V = [v_{ij}]$ of the connectivity matrix **A**, where $v_{ij}$ measures the controllability of mode $\lambda_j(A)$ from control node *i*. We can then define a scaled measure of the controllability of all N modes, $\lambda_1(A), \ldots, \lambda_N(A)$, from neuron *i* as:

$$\phi_i = \sum_{j=1}^{N} \left(1 - \lambda_j^2(A)\right) v_{ij}^2, \qquad [6]$$

We assign each neuron a ranked value between 1 and N based on their ϕ value, with 1 being the neuron with the lowest modal controllability (lowest ϕ) and N being the neuron with the highest modal controllability (highest ϕ).

**Neuronal Activity & Network-Wide Bursts**

Now that we have defined strategies to target the addition of autapses to neurons, we wish to understand their role in controlling global network dynamics. We therefore define several summary statistics of neural dynamics including firing rate and network-wide bursts, which are coordinated firing events across large numbers of neurons within a brief time period.

Note that other complementary definitions of what consists a network-wide burst can be found in the literature, and we briefly review them in the Supplement.

Quantitatively, we define network-wide bursts as periods of activity in which the number of neurons firing at the same time met or exceeded a threshold level of 40% of neurons in a millisecond. We implemented a 5ms tolerance in the burst detection algorithm, combining two groups of neurons into a single burst if they fired within 5 ms of each other. This burst detection algorithm was robust to changes in the threshold level of neurons that must be active for a period of activity to be considered a burst (Supp. Fig. 1).

After defining bursts, we calculated the mean and standard deviation of three summary statistics for each simulation: the burst frequency, the interburst-interval, and the burst duration. We defined the burst frequency to be the mean number of bursts per second across the simulation. We calculated the mean interburst-interval (IBI) from the temporal midpoints of the bursts in each simulation. Finally, we defined the mean burst duration as the fraction of simulation time spent in the network-wide bursting state. To summarize these results across simulations, we calculated either an unweighted mean or standard deviation (for burst frequency) or a weighted mean or standard deviation (for burst duration and average IBI). We used a weighted mean for the second two statistics because each simulation was built on a different connectivity matrix, and therefore could display different bursting parameters. For the IBI and burst duration, we computed the weighted average, $X_{weighted}$, as follows:

$$X_{weighted} = \frac{\sum_{i=1}^{n} w_i x_i}{\sum_{i=1}^{n} w_i}, \qquad [7]$$

where $x_i$ is the data value (the average IBI or burst duration from the individual simulation), the weight $w_i = 1/\sigma_i^2$, $\sigma_i$ is the standard deviation of $x_i$, and $n$ is the number of data values (number of independent simulations). We also calculated the weighted standard deviation

$$\sigma_{weighted} = \frac{1}{\sqrt{\sum_{i=1}^{n} w_i}}, \qquad [8]$$

for the IBI and burst duration across simulations. Together, these parameters described the bursting behavior of the networks with and without autapses.

**Statistical Analysis**

We used JMP Pro 11 (SAS Institute Inc.) for all statistical analyses. To identify autaptic conditions where the addition of autapses induced significant changes in burst frequency, we performed a mixed-model ANOVA using the Full Factorial Repeated Measures ANOVA Add-In (https://community.jmp.com/docs/DOC-6993, Julian Harris, SAS employee). Each original, non-autaptic connectivity matrix was treated as a "subject." The between-subjects factor was the type of connection added (i.e., autapses, non-autaptic inputs, or non-autaptic outputs) while the within-subjects factor was the percent connections added, either autaptic or non-autaptic connections. When the interaction effect between the type of connection added and the percent connections added was significant, we performed post-hoc analyses using Tukey's HSD test. These post-hoc results were used to identify targeting strategies and simulation parameters in which adding autapses significantly changed burst frequency from that of the original non-autaptic network and from the appropriate null models. Separate statistical analyses were completed for each of the seven targeting strategies and for each possible pair of excitatory and inhibitory strength values.

  To identify differences in burst frequency induced by adding connections according to the seven targeting strategies, we again performed a mixed-model ANOVA using the Full Factorial Repeated Measures ANOVA Add-In (https://community.jmp.com/docs/DOC-6993, Julian Harris, SAS employee) where each original connectivity matrix was a "subject." Here, the within-subjects factor was again the percent connections added. The between-subjects factor was the targeting strategy used to select neurons to which to add autaptic or non-autaptic connections. When there were significant effects of the interaction between the type of connection added and the targeting strategy, we performed post-hoc analyses using Tukey's HSD test. These post-hoc results were used to identify the simulation parameters between which the burst frequency significantly differed. Additionally, when there were significant effects of the targeting strategy, we performed post-hoc analyses to identify which targeting strategies were significantly different from one another. In the main text, we report results comparing three targeting strategies: neurons chosen by highest average controllability, highest modal controllability, and uniformly at random. In the supplement, we show results for all seven targeting strategies (see Supp. Fig. 2.). Separate statistical analyses were performed for each type of connection added within each excitatory/inhibitory strength combination.

# Results

## Network Construction

We modeled networks of cortical neurons by placing excitatory and inhibitory neurons on the surface of a sphere (Fig. 1a) and connecting them – with no autapses – in a distance-dependent manner (Fig. 1b). We next modified these networks by adding autapses (Fig. 1c), or self-loops, to groups of excitatory or inhibitory neurons chosen according to certain characteristics of their local network neighborhood. We also constructed null model networks by adding input or output connections on the selected group of neurons to account for the increase in the number of connections in a network in which autapses were added (Fig. 1c). These null models were used to test whether the observed dynamical changes were simply due to the increase in connections in the autaptic networks or were more interestingly due to the autaptic nature of the connections specifically. We study the effect of autapses on several summary statistics of bursting behavior (Fig. 1d). See Materials and Methods for additional details.

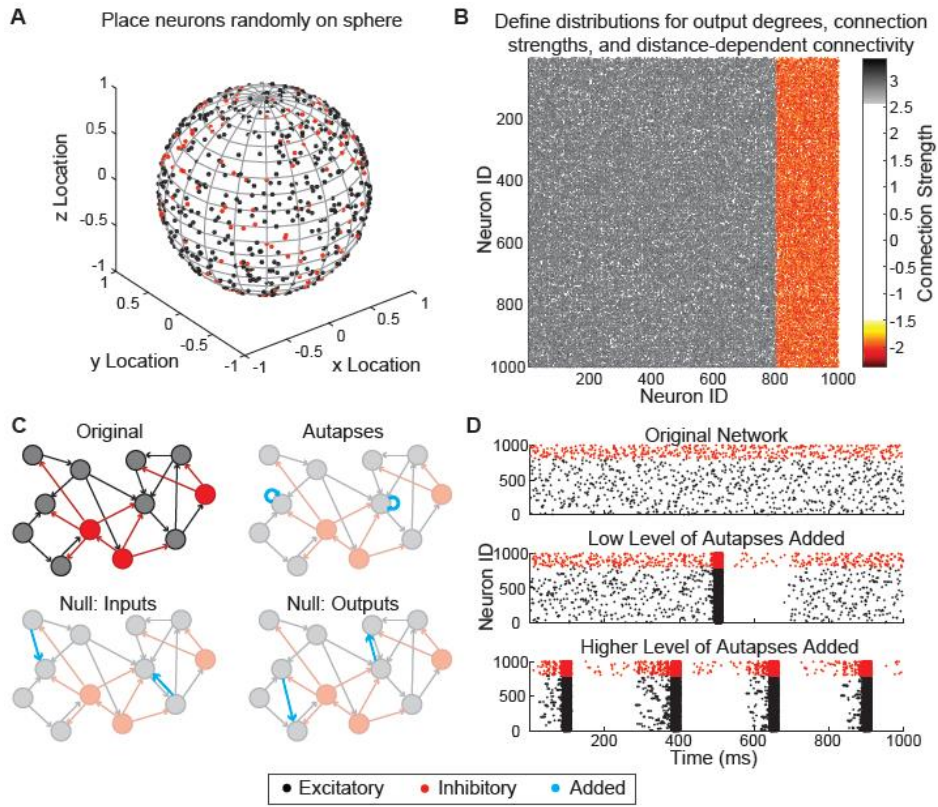

**Figure 1. Schematic of Empirical Methods**. *(A)* Excitatory and inhibitory neurons were placed uniformly at random on the surface of a sphere. *(B)* Coupling between neurons was constructed using well-defined distributions for node strength, and by implementing a set probability of connection fall-off with distance. The resulting coupling matrix is a weighted, directed adjacency matrix. *(C)* Autapses were added to either inhibitory or excitatory neurons to determine their effect on network dynamics, in comparison to appropriate statistical null models where input or output connections were added in place of autapses. *(D)* We measured the effects of autapses on the dynamics of the network activity by quantifying firing rate and network-wide bursts.

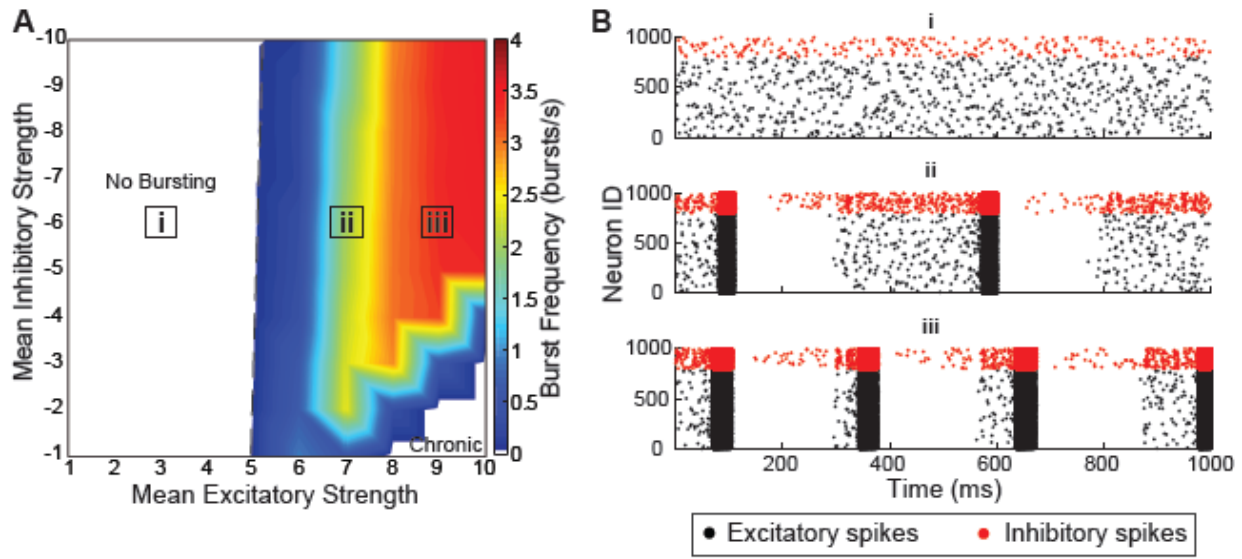

**Figure 2. Dynamics of neuronal network simulations without autapses.** *(A)* Burst frequencies at different excitatory and inhibitory strengths, showing regimes of no bursting [i], intermittent bursting [ii], and chronic bursting [iii]. *(B)* Example raster plots displaying activity at three different excitation levels.

**Network Dynamics**

We observed no change in dynamics if we simulated from 120 seconds to 3600 seconds of neural activity in the network. Therefore, we simulated 120s of activity for networks with different levels of mean excitatory and inhibitory strength (Fig. 2). Based on the amount of excitation and inhibition present in the network, we identified three distinct regimes of neural dynamics. At low excitation levels, independent of inhibition level simulated, network-wide bursting never occurred (Figure 2B, i). Activity in this regime was asynchronous and dominated by noise. As the excitation level increased, activity became less dependent on noise and network-wide bursts occurred more frequently and more regularly (Figure 2B, ii). At high excitation levels, the network entered a chronic bursting regime (Figure 2B, iii). The decrease in burst frequency at very high excitatory and very low inhibitory strengths is due to this transition to chronic bursting; burst frequency is decreasing as burst duration is increasing (see Supp. Fig. 3).

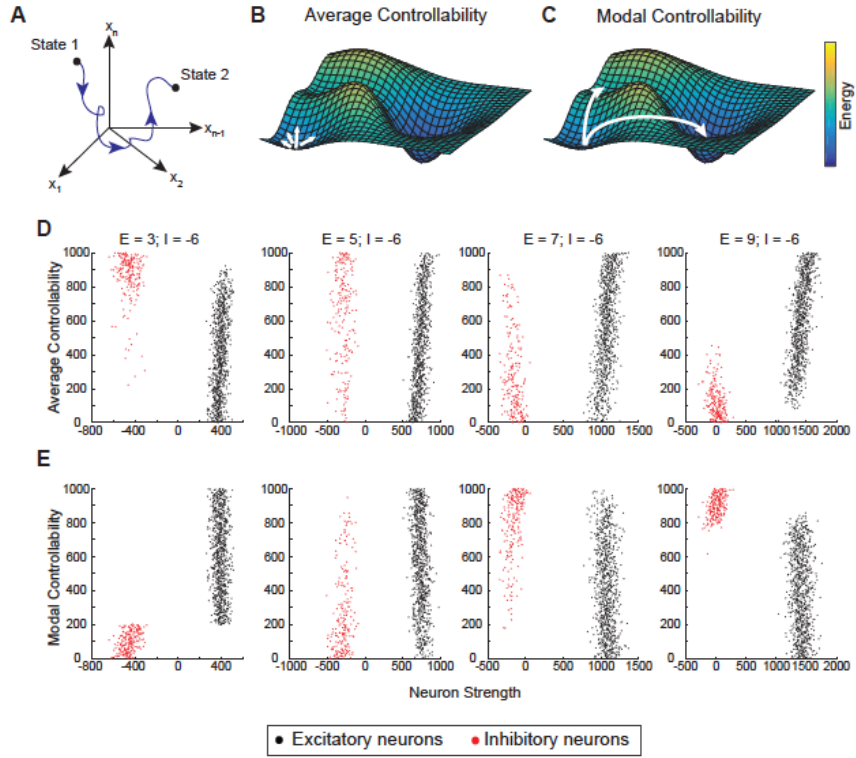

**Figure 3. Notions of Network Control and Their Relationships to Topology.** *(A)* Schematic illustrating the idea that a controllable network can be driven from an initial state to a final state in some multidimensional landscape within a finite time period. *(B)* Illustration of a 3-dimensional energy landscape on which nodes with high average controllability drive the system from a baseline state to many easily reachable states (arrows). *(C)* Illustration of the same 3-dimensional energy landscape on which nodes with high modal controllability drive the system to difficult-to-reach states, from one energy minimum to another over a large energy barrier (arrows). *(D)* As excitatory synaptic strength in the network is increased while inhibitory strength is held constant, the nodes with the highest average controllability shift from the inhibitory to the excitatory network. *(E)* Over the same span of network configurations, highest levels of modal controllability appear in the excitatory network at low synaptic strength and shifts to the inhibitory network at high excitatory synaptic strengths..

**Location of Control Points in the Network Depends on the Levels of Excitation and Inhibition**

Controllability describes the potential to drive a dynamical system from an initial state to a desired final state given that inputs are applied to one or more nodes in the network (Fig. 3a). The importance of individual nodes in driving the system to certain states can be quantified using distinct control strategies. Two commonly studied control strategies are average control and modal control. Nodes with high average controllability are theoretically predicted (by a simplified model of linear system dynamics) to drive the system to many energetically easy-to-reach states (Fig. 3b). Nodes with high modal controllability are theoretically predicted (again, by a simplified model of linear system dynamics) to drive the system to many difficult-to-reach states (Fig. 3c).

In the simulated networks, we found that average and modal controllability were directly related to node strength in excitatory neurons (Fig. 3d-e). In the excitatory population, we observe a positive correlation between average controllability and the sum of the inputs onto a neuron, and a negative correlation between modal controllability and the sum of the inputs onto a neuron. These relationships are consistent with those observed in undirected networks representing large-scale white matter connectivity in the human brain [16, 52].

Interestingly, the relationship between controllability and strength in the inhibitory neurons was less clear, as we observed a non-trivial dependence between the controllability statistics and the balance between excitation and inhibition in the network. Specifically, when the mean inhibitory strength was larger than the mean excitatory strength, excitatory neurons displayed lower average and higher modal controllability values, while inhibitory neurons displayed higher average and lower modal controllability values. In contrast, when the excitatory strength was larger than the inhibitory strength, excitatory neurons tended to display higher average and lower modal controllability values than inhibitory neurons. As the difference between the magnitudes of excitatory and inhibitory strength increased, the separation between controllability values of the excitatory and inhibitory neurons became more prevalent.

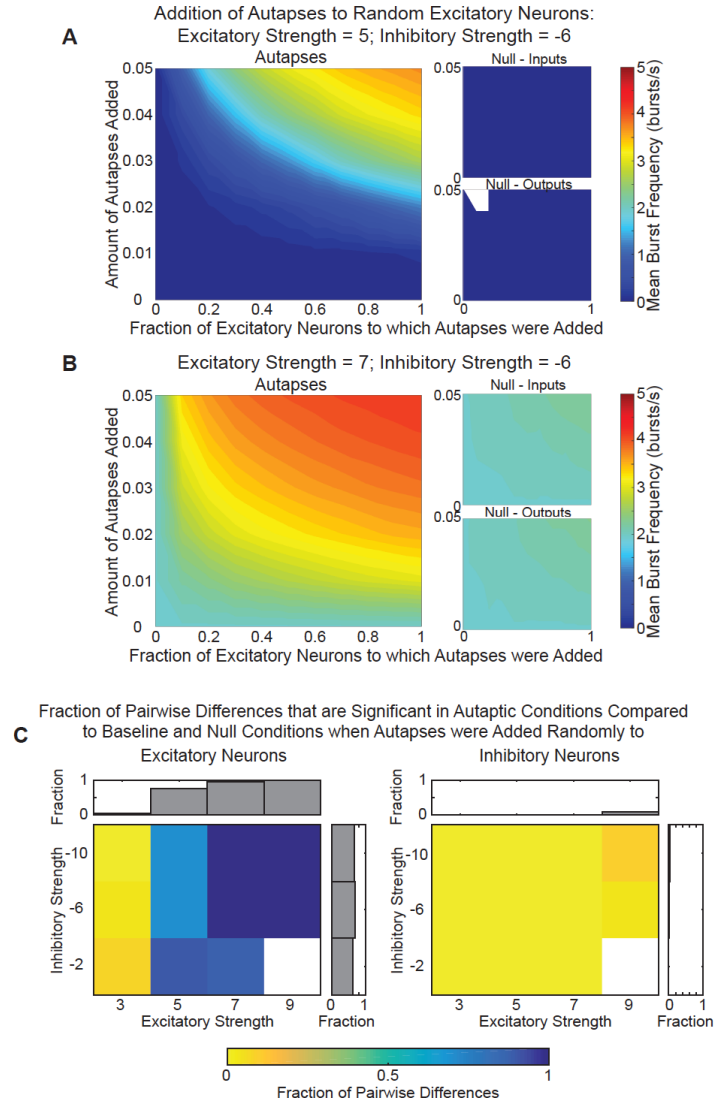

**Figure 4. Role of Autapses on Excitatory Neurons in Network-Wide Bursts.** *(A)* Burst frequency for an example simulation in the autaptic model (left) in which autaptic connections were added to excitatory neurons uniformly at random, or in two null models (right) in which non-autaptic connections were added to maintain either the number of input connections (an input null model) or to maintain the number of output connections (an output null model). This network used a mean excitatory and inhibitory strength of 5 and -6, respectively. The abscissa shows the fraction of the excitatory neurons that had modified connectivity. The ordinate gives the amount of connections added, defined as a fraction of the neuron's original number of outputs. *(B)* The same type of information presented in panel (A), except here shown for a network that had a mean excitatory strength of 7 and a mean inhibitory strength of -6. (C) The fraction of autaptic conditions (fraction of neurons x amount of autapses; see Methods) within a given excitation and inhibition level that were significantly different from the original baseline network and from the corresponding input and output null models when autapses are added to excitatory (left) *versus* inhibitory (right) neurons. Bar graphs show the fraction of conditions that were significantly different from the baseline network and from control conditions across a particular excitatory or inhibition level.

**Adding Autapses to Excitatory Neurons Increases Burst Frequency**

We added varying amounts of autapses or non-autaptic connections to a specified fraction of randomly selected excitatory or inhibitory neurons throughout the network (Fig. 4). At lower excitatory strengths, burst frequency increased with both the percent of autaptic neurons and the number of autapses added to autaptic neurons. However, adding non-autaptic connections in the null model simulations did not increase burst frequency (Fig. 4a). At higher excitatory strengths, burst frequency again increased with the percent of autaptic neurons and the number of autapses added (Fig. 4b). Here, unlike at lower excitation levels, adding non-autaptic connections in the null model simulations also increased bursting frequency; however, higher levels of bursting still occurred in the autaptic network compared to the null model networks.

We can summarize the data described above by computing the fraction of autaptic conditions (fraction of neurons x amount of autapses; see Methods) within each excitation and inhibition combination that had significantly different burst frequencies from those observed in the original non-autaptic system and from the burst frequencies of the corresponding null model systems (Fig. 4c). We observed that adding autapses to excitatory neurons induces greater changes in burst frequency than adding autapses to inhibitory neurons. Moreover, we observed that, when adding autapses to excitatory neurons, the level of excitation more strongly affects changes in the network's bursting behavior than the level of inhibition.

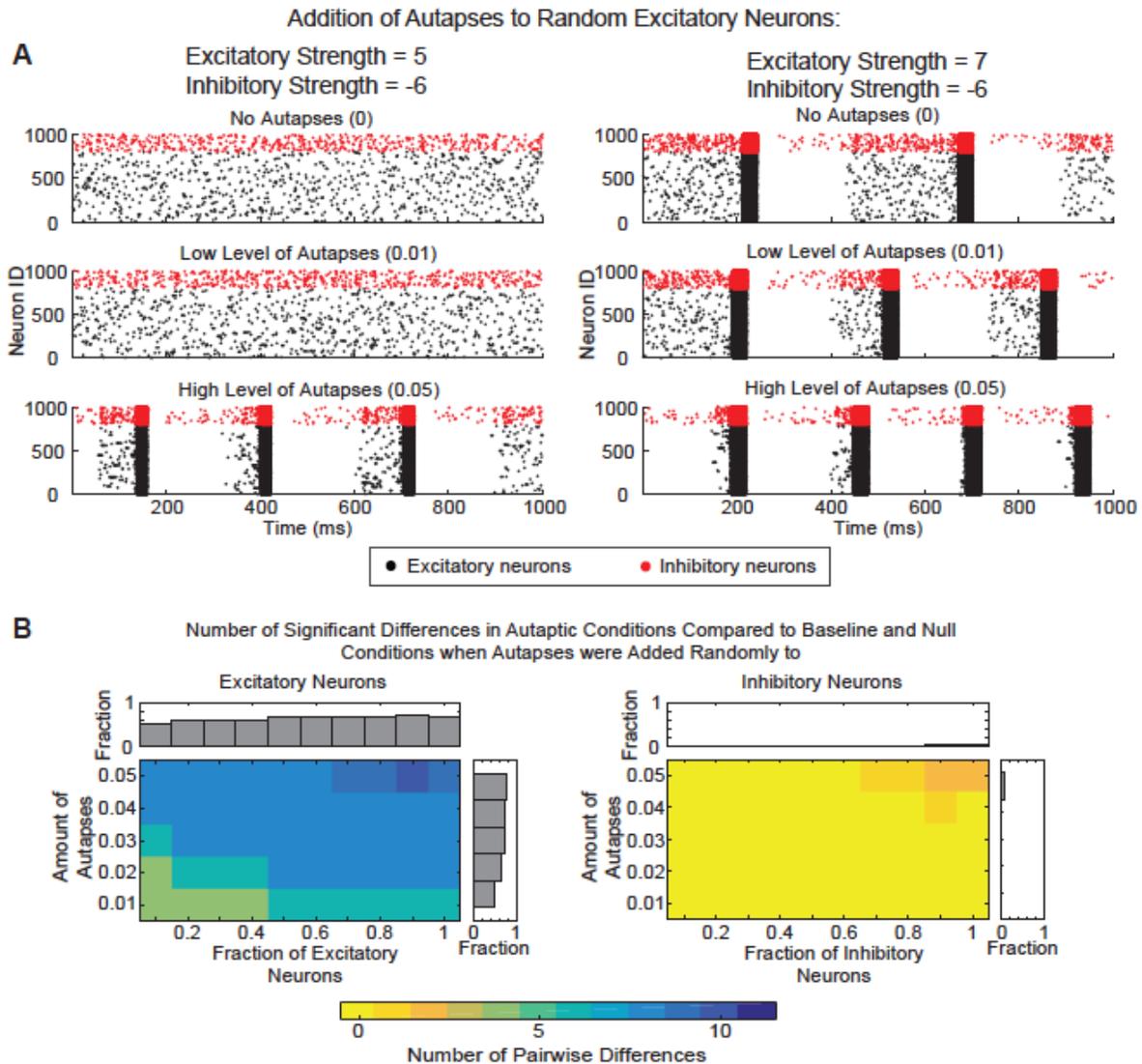

**Figure 5. Effect of Amount of Autapses on Network Dynamics.** *(A)* Raster plots showing 1s of activity for networks with lower *(left)* and higher *(right)* excitation levels, as the amount of autpases is varied from 0% *(top)* to 5% *(bottom)*. *(B)* The fraction of excitatory and inhibitory strength pairs at which the bursting frequency after addition of autapses was significantly different from the bursting frequency at baseline or in the null models. Number of pairwise differences are given as a function of the amount of autapses added, as well as the fraction of excitatory (left) and inhibitory (right) neurons. Bar graphs show the fraction of excitatory/inhibitory strength pairs that produced bursting frequencies that were significantly different from the baseline network and control conditions across the amount of autapses added.

Next we asked how these results depended on the number of autaptic connections that were added to the network. Networks constructed with more autapses displayed an increased burst frequency and burst regularity compared to networks constructed with fewer autapses (Fig.

5a). Interestingly, the relationship between burst frequency and number of autapses was modulated by the mean excitation strength of the system. At lower excitation levels, more autapses were needed to induce the same differences in burst frequency observed at higher excitation levels.

Importantly, these results do not address the question of whether the important driver of bursting dynamics is simply the number of autaptic connections, or whether the more fundamental parameter is the number of autapses per neuron. To directly address this question, we computed, for each autaptic condition (fraction of neurons x amount of autapses), the fraction of the excitatory/inhibitory strength combinations with burst frequencies that were significantly different from baseline and from the input and output null models. Again, we observed larger fraction of differences in bursting dynamics due to the addition of autapses to excitatory rather than to inhibitory neurons (Fig. 5b). We also observed that the amount of autapses added to an excitatory neuron played a larger role in driving the increase in burst frequency than the fraction of excitatory neurons in the network that were autaptic (Fig. 5b, bar graphs). These results demonstrate the effect of autapses on network dynamics is nonlinear because the same number of autaptic connections added to fewer (more) neurons has a greater (lesser) impact on burst frequency.

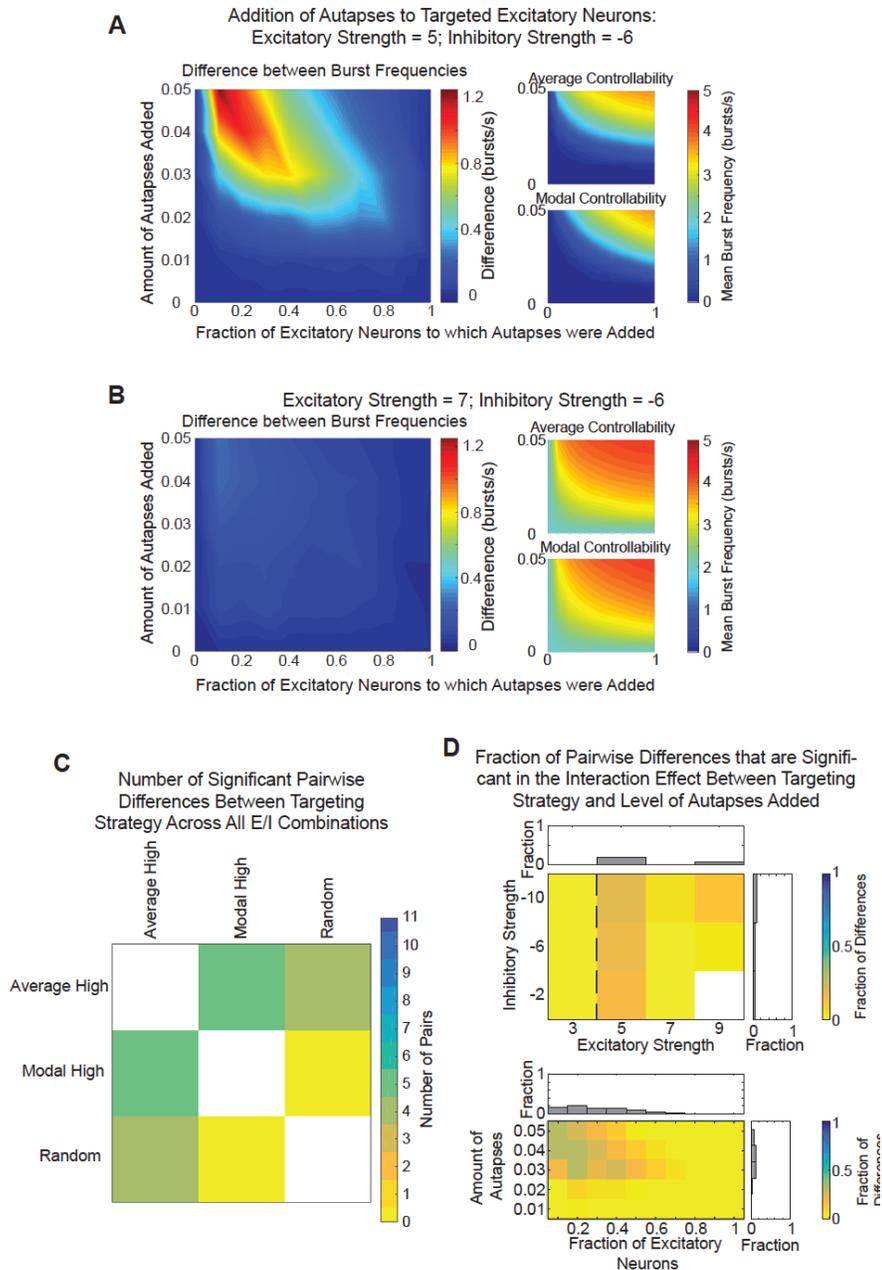

**Figure 6. Targeting Autapses to Control Points in the Network.** *(A, left)* Difference in the burst frequency when autapses were added to the highest average controllability neurons versus the highest modal controllability neurons, as a function of the fraction of excitatory neurons to which autapses were added, for example simulations. *(A, right)* Observed burst frequency when autapses were added to the highest average (top) versus highest modal (bottom) controllability neurons for example simulations. The abscissa shows the fraction of the excitatory neurons to which autapses were added. The ordinate gives the amount of connections added, defined as a fraction of the neuron's original number of outputs. *(B)* Similar data to that presented in panel *(A)* except here for a higher level of excitation. *(C)* Number of significant pairwise differences – across all eleven excitation and inhibition combinations – in burst frequency for networks constructed from the two targeting strategies. *(D, top)* Fraction of significant pairwise differences

in burst frequency in the interaction effect between targeting strategy and amount of autapses added, within each excitatory/inhibitory strength combination. *(D, bottom)* Fraction of pairwise differences in burst frequency across excitatory and inhibitory strength combinations at each autaptic condition (fraction of neurons x amount of autapses). Bar graphs show the fraction of excitatory/inhibitory strength pairs that produced bursting frequencies that were significantly different between the targeting strategies across the amount of autapses added.

**Targeting Autapses to Control Neurons Differentially Impacts Burst Frequency**

After studying the effect of autapses added to neurons chosen uniformly at random, we next asked whether we could target autapses to specific "control" neurons to increase burst frequency even further. To address this question, we examine bursting dynamics when autapses are added to either excitatory or inhibitory neurons with either the highest average or highest modal controllability values (for definitions, see Methods). At a mean excitatory strength of 5, adding autapses to excitatory neurons with the highest average controllability resulted in higher burst frequencies at certain autaptic conditions than when autapses were added according to the highest modal controllability (Fig. 6a). At a stronger excitatory level of 7, although there was a significant interaction effect between the amount of connections added and the targeting strategy, we did not observe significant differences between corresponding autapse levels of average and modal controllability (Fig. 6b).

To better understand the impact of excitatory and inhibitory strength values on these results, we calculated the number of significant differences in burst frequency between average and modal controllability targeting strategies that occurred across the 11 excitatory and inhibitory strength combinations (Fig. 6c). We observed a maximum of 5 differences, which can be explained by Fig. 5d where we see that significant pairwise differences between bursting dynamics observed in different targeting strategies only occurred at excitation/inhibition levels of 5/-2, 5/-6, 5/-10, 7/-10 and 9/-10. Additionally, the majority of significant differences between targeting strategies occurred when autapses were added to less than half of the excitatory neurons (Fig. 6d). As we added autapses to increasingly more neurons according to these two opposing targeting strategies, the overlap between the groups of neurons that targeting strategies selected increased, making the rules more similar and leading to fewer observed differences in bursting dynamics.

Results from all targeting strategies are shown in Supplementary Information. No targeting strategy or interaction effect was observed when we added autapses to inhibitory neurons.

## Discussion

Here we examine the relationship between theoretical measures of structural controllability and observed measures of network dynamics. We build on a well-developed numerical simulation of cortical and hippocampal network dynamics to study the influence of autaptic connections on bursting frequency and regularity. Autaptic connections are represented as self loops in the network and present unique control features whose impact on neuronal network dynamics is unknown. We show that these self-loops differentially influence network dynamics: when applied to excitatory (but not inhibitory) neurons, these self-loops lower the threshold for network bursting. Directing self-loops to nodes of high average controllability, which are theoretically predicted to effectively move the system into local easily-reachable states, leads to an increase in the frequency and regularity of network-wide bursts. Together, these results suggest a role of autaptic connections in controlling network-wide bursts in diverse cortical and subcortical regions of mammalian brain.

**Dynamic Behaviors Driven by Structural Network Architecture.** In our network, the most salient outputs are the appearance of bursting or synchronization of the network, and the corresponding interburst intervals, that appear over time. Synchronization of brain networks is often considered to be key for learning [56], memory [57, 58], and other higher-order cognitive processes [59-61]. In contrast, sporadic or sustained bursting can lead to the development of pathological networks in diseases such as epilepsy [62]. Our results describing a broad class of bursting types are consistent with previous models showing a dynamic range of activity in neuronal systems, including the coherent activity observed in health and the abnormal activity observed in disease. These results further add to the literature by demonstrating that the observed dynamics (burst frequency and regularity) are directly driven by the underlying network connectivity and synaptic weights between neurons. These networks were designed to model only local microcircuit architectures with no delay among neurons in the network. Our findings provide insight into how these local self-loops can regulate the neural dynamics of these microcircuits. A key observation is that spreading autaptic connections among a number of excitatory neurons affected the output neural dynamics (bursting) more significantly than concentrating many autapses to a smaller number of neurons. From a network perspective, this general result indicates that drivers of network behavior exist preferentially at the level of single nodes, rather than at the level of single edges within the network.

**Autapses as Effective Drivers of Shifting Network Dynamics.** Physiological estimates of autaptic connections in excitatory neurons rarely exceed 1-2% of neurons within a network [refs 1-7], while some interneurons can display significant levels of self-inhibition [refs 5-7]. It is interesting to note that we observed significant transitions network-wide behavior when our simulations extended beyond these physiological conditions. These data support the plausible intuition that neuronal networks *in vivo* operate at an optimal point for shifting network dynamics by the deletion or the addition of only a few self-loops, supporting maximal flexibility or dynamic range. This type of self-loop modulation might occur as a function of synaptic pruning that is common during neuronal development, which can work to consolidate network dynamics towards a stable equilibrium point [63]. An alternative potential mechanism for self-loop formation is sprouting, commonly observed after injury, which could transform a low activity network into a highly active network with periods of synchronization [64]. The impact of these dynamics are less clear and depend on the frequency of the neuronal activity, with an enhancement of activity potentially promoting prosurvival signaling through the nuclear activation of antioxidant signaling pathways [65, 66]. Alternatively, extensive aberrant sprouting could drive the network into a state of overexcitation, which could in turn lead to targeted neurodegeneration from chronic, seizure-like bursting of the network.

**Linear Predictions of Nonlinear Dynamics.** A key question we explored was how the nonlinear dynamics of this commonly studied network were influenced by the patterns of structural connections surrounding single neurons. To gain an understanding of this relationship, we drew from the field of structural controllability [20-22]: a subfield of control and dynamical systems theory that offers predictions of which nodes in a network might act as control points under the assumptions of a simplified linear dynamics. We asked whether these predictions offered fundamental utility in understanding the complex behaviors of neuronal networks. We observed that for a range of excitatory and inhibitory synaptic strengths, the structural network change elicited by adding autapses to putative control points in the network increased burst frequency and regularity, to a much greater degree than adding autapses to neurons chosen uniformly at random. These results demonstrate that the predictions of control points derived from a simplified linear model of neuronal network dynamics are supported by observed changes in network dynamics, consistent with reported results at larger spatial scales [53]. It will be interesting in future to study the role of alternative control strategies (including boundary controllability; [16, 18]) in the other areas of the excitation/inhibition phase space characterized by other network behaviors including either continuous bursting or the lack of bursting.

**Inference of network dynamics through controllability.** Although we observed correspondence between measures of controllability and the resulting dynamical behavior of networks, we recognize the activation and bursting phenomenon that appears in the networks is a nonlinear process. As the linear approximation of inherent nonlinear systems is under active study, we found many corresponding connections between linear control theory and bursting behavior, but these were not complete. For example, moving a network that is already bursting into a higher bursting state may be viewed as an easily approachable new state, and control theory would predict nodes with high average controllability would be ideal for moving the network into this new state. This is consistent with our observations, as excitatory neurons represented neurons of high average controllability in these networks and adding autaptic connections specifically to these neurons moved the network into a higher bursting state. Likewise, a network that is currently not bursting can reach another easily reachable non-bursting state by adding autaptic connections to inhibitory neurons, as these neurons represent a majority of the neurons with high average controllability. In comparison, shifting a network into a difficult to reach state – e.g., moving a bursting network into a non-bursting network – suggests that neurons with high modal controllability would be the likely targets. However, in such a network, the inhibitory neurons represented nearly all of the nodes with high modal controllability and adding autaptic connections did little to affect the dynamics. Although this may highlight one gap in using linear control theory to predict nonlinear dynamics of neural networks, it is worth noting that we could shift bursting networks into nonbursting networks at higher levels of inhibitory synaptic strength, suggesting at least a regime of the network where the predictions align across the two domains.

**Methodological Considerations.** There are several important limitations to this work that could be explored in future studies. First, these simulations do not provide more detailed mechanisms of network remodeling (e.g., spike timing dependent plasticity, homeostatic plasticity, presynaptic facilitation) that may affect the temporal evolution of bursting that can occur by including more detailed models of synaptic currents. However, we do not anticipate that these more detailed features of the model would affect our general result of changes in bursting dynamics associated with changes in underlying network architectures. Second, we also examined a range of excitatory and inhibitory synaptic strengths used in past studies, and found that the principal change in dynamical behavior was mediated through the excitatory neurons. Extending the current work into regimes where the inhibitory synaptic strength also influences the bursting behavior of the network would test the robustness of our observations for network controllability where alterations in either excitatory or inhibitory strength would lead to changes in the neural dynamics. Third, we directly addressed the question of whether predictions of control points in the network derived from a simplified linear model of neuronal dynamics could be used to understand the nonlinear dynamics of the full model. It would be interesting in future to develop and apply techniques from nonlinear control theory to further understand the mapping between control points and observed network dynamics. Indeed, studying nonlinear control strategies could offer particular utility in extending these examinations to the study of the switch timing of transcriptional activators and repressors within genetic circuits that code for neuronal function. Finally, the notions of average and modal controllability are agnostic to the specific initial and final states of the system, offering predictions based on the ensemble of local easily-reachable states (average controllability), and the ensemble of distance difficult-to-reach states (modal controllability). It could be interesting in future work to study specific transitions of the neuronal network from a specified initial state of activation to a specified final state of activation, potentially offering insights into the finite set of transitions that a network is expected to display under normal operating conditions [52, 67].

**Conclusion.** In this study, we focus predominantly on descriptive statistics and simple predictors of future network performance and dynamic behavior. However, the time is ripe for the field of network neuroscience to take the next step in the *de novo* design of networks theoretically optimized for specific types of computations. These design efforts capitalizing on generative modeling frameworks would be especially important for understanding the different structure-function mappings observed across different regions of cortex, as well as in health *versus* disease, and to posit therapeutic interventions for network reconfiguration and recovery. We anticipate that the targeted placement of autaptic connections will be an important dimension of these solutions, as well as more generally being critical for our understanding of the dynamics observed in translation, transcription, and gene regulatory networks.

**Additional information**

**Author contributions:** LW completed the simulations, analyzed the data, and participated in writing the manuscript and constructing the figures. SG and FP contributed algorithms for calculating controllability and provided suggestions for figure development. DSB and DM planned the study, participated in the analysis and interpretation, contributed to the writing and revising of the manuscript, and contributed to the figure development. All authors reviewed the manuscript.

**Acknowledgments:** Funding was provided by U.S. Department of Health & Human Services (NIH NS088176, NS093293) and the Army Research Office (ARO) - W911NF-10-1-0526.

**Competing financial interests:** The author(s) declare no competing financial interests.

# Supplementary Materials

**Section 1: Supplementary Results**

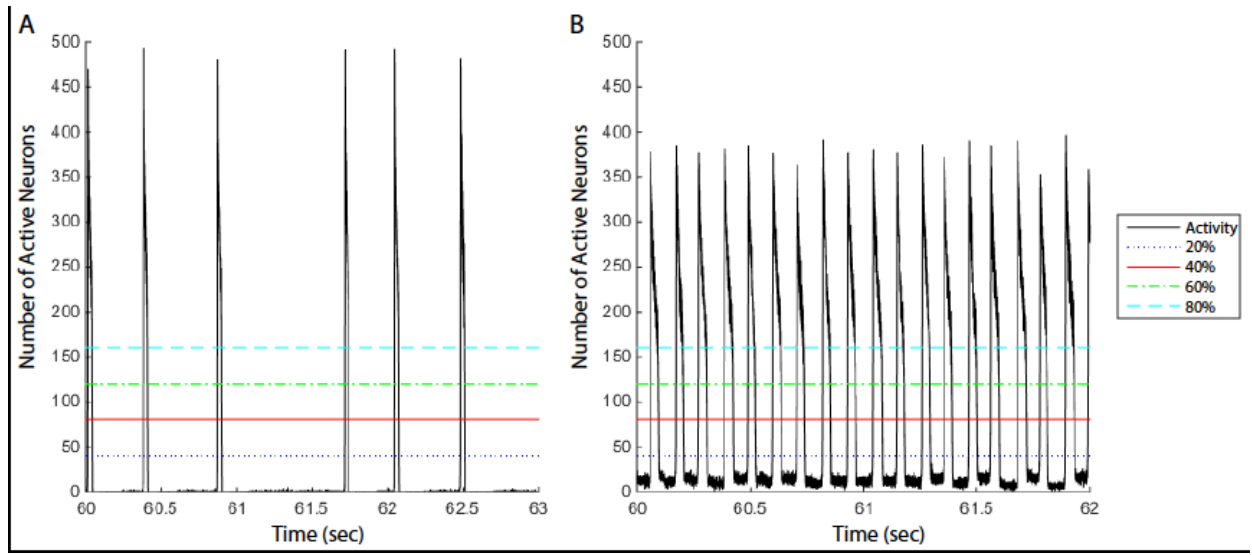

**Supplementary Figure 1. Robustness of results to burst detection threshold.** In the main manuscript, we define network-wide bursts as periods of activity in which the number of neurons firing at the same time met or exceeded a threshold level of 40% of neurons in the network divided by the number of steps per ms. Here we show that our results were robust to changes in the threshold level of neurons that must be active for a period of activity to be considered a burst. (A) Number of active neurons as a function of time in seconds for an example simulation at the excitatory strength of 7. (B) Number of active neurons as a function of time in seconds for an example simulation at the excitatory strength of 9. Colored horizontal lines indicate different threshold choices (20%, dark blue dotted line; 40%, red solid line; 60% green dot-dashed line; 80%, cyan dashed line). Note that the number of network-wide bursts is identical across these threshold choices.

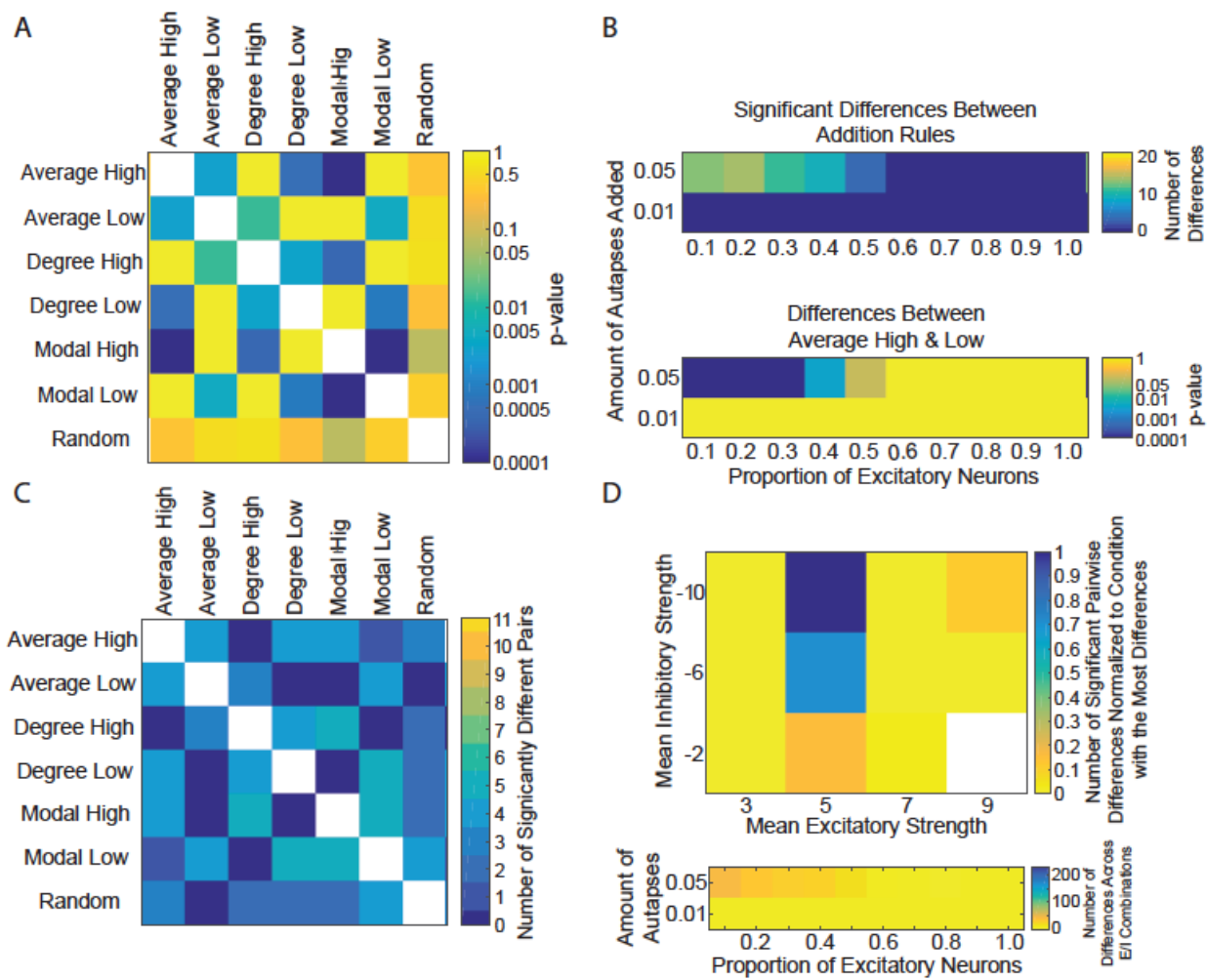

**Supplementary Figure 2. Effect of targeting strategy.** In the main text, we report results comparing three targeting strategies using a mixed-model ANOVA: neurons chosen by highest average controllability, highest modal controllability, and uniformly at random. Here, we use the same statistical procedure to show results for all seven targeting strategies. (A) Differences in burst frequency induced by adding connections according to the seven targeting strategies, Color indicates p-values of posthoc Tukey's HSD test. (B) Dependence of results in (A) as a function of both the proportion of excitatory neurons (x-axis) and the amount of autapses added (y-axis). The number of significant differences between addition rules was greater for simulations with more autapses and smaller proportion of excitatory neurons. (C) Number of significantly different pairs of simulations (across different levels of excitatory and inhibitory strength, and across different amounts of autapses) for each set of targeting strategies. Color indicates number

of significantly different pairs. (D) Dependence of summary results in panel (C) on the inhibitory/excitatory strength combination (top) and the amount of autapses added (bottom).

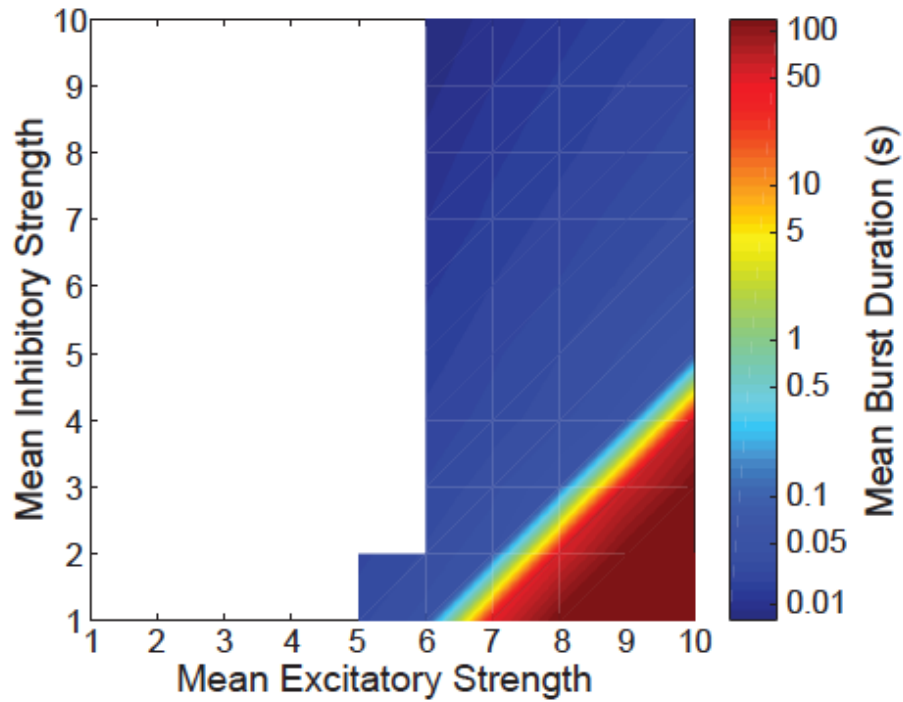

**Supplementary Figure 3. Burst duration dependent on excitatory and inhibitory strength.** Mean burst duration (color) as a function of mean excitatory strength (x-axis) and mean inhibitory strength (y-axis).

### Section 2: Supplementary Discussion

In the main manuscript, we utilize a mathematical definition of a network-wide bursts. However, it is important to note that there is no one well-accepted definition of a network-wide burst. To illustrate this point, here we describe previously-used definitions, which vary both quantitatively and qualitatively.

1. In a previous computational model, the resulting spiking activity was binned into 10 ms time windows. A burst was identified when more than 25% of the neuronal population fired during that time window [1].

2. In a different study that included both a computational model and multi-unit activity (MUA) from multielectrode array recordings of *in vitro* rat cortical neurons, a "network spike" was detected by counting the spikes recorded at all electrodes in 50 ms time windows. If the number of spikes detected exceeded a threshold of 25% of the maximum spike count recorded, then a burst was detected [2].

3. In another analysis of MUA activity, recordings from each electrode were searched for sequences of sequences of at least four spikes with inter-spike intervals less than a threshold set to the less of 100 ms or ¼ of that electrode's inverse spike detection rate. If a group of these sequences overlapped in time across multiple electrodes, it was called a burst [3].

4. Pasquale et al [4] developed a burst detection algorithm involving two cases based on the separation of the two peaks (bursting ISI peak and non-bursting ISI peak) logarithmic histogram of the ISI. If the minimum between the two peaks is less than 100 ms, that minimum is used as the maximum ISI allowed within a burst and used to identify spikes within a burst. On the other hand, if the minimum is greater than 100 ms, a 100 ms threshold is used to detect burst cores and the location of the minimum between the peaks is used as the absolute bursting threshold. Spikes occurring within the burst core ISI threshold are identified as burst core spikes while spikes occurring at rates between the burst core ISI threshold and the minimum on the histogram are referred to as burst boundary spikes.

5. Recent work utilized and compared the results of several burst detection algorithms. These included: (a) a rate-threshold detector in which a burst was detected if a firing rate histogram with 50 ms time windows exceeded a threshold number of spikes, (b) a ISI-threshold detector in which interspike interval thresholds were set at the minimum between peaks of the logarithmic ISI probability distribution and capped at 10 ms. If the ISIs of five consecutive spikes were each less than the ISI threshold, then a single-channel burst was detected. A network burst required 20% of the recording channels to be activated. (c) a rank surprise detector in which the minimum number of spikes and maximum ISI within a burst were 5 spikes and 100 ms, respectively, and at least 10 single channel bursts overlapped in time, and (d) an $ISI_N$-threshold method in which the user chooses a minimum network burst size in terms of N, the number of spikes that make the smallest network burst, and then if N consecutive spikes occur within a time period less than or equal to the $ISI_N$-threshold, where $ISI_N$ is the ISI between every Nth spike in the network and the

$ISI_N$-threshold is the minimum between regions of bursting and non-bursting in a logarithmic histogram of $ISI_N$, the time period is a burst [5].